\documentclass{article}
\usepackage[accepted]{ml4astro2025}
\usepackage{microtype}
\usepackage{graphicx}
\DeclareGraphicsExtensions{.pdf,.png,.jpg} 
\usepackage{subfigure}
\usepackage{booktabs} 

\usepackage{hyperref}

\usepackage{amsmath}
\usepackage{amssymb}
\usepackage{mathtools}
\usepackage{amsthm}

\usepackage[capitalize,noabbrev]{cleveref}

\theoremstyle{plain}

\theoremstyle{definition}

\theoremstyle{remark}

\usepackage[textsize=tiny]{todonotes}

\mlforastrotitlerunning{Degeneracy-Aware Pulsar Parameter Estimation from Light Curves via Deep Learning and Test-Time Optimization}

\begin{document}
\twocolumn[
\mlforastrotitle{Degeneracy-Aware Pulsar Parameter Estimation from Light Curves via Deep Learning and Test-Time Optimization}



\mlforastrosetsymbol{equal}{*}

\begin{mlforastroauthorlist}
\mlforastroauthor{Abu Bucker Siddik}{equal,yyy}
\mlforastroauthor{Diane Oyen}{yyy}
\mlforastroauthor{Soumi De}{yyy}
\mlforastroauthor{Greg Olmschenk}{comp}
\mlforastroauthor{Constantinos Kalapotharakos}{comp}

\end{mlforastroauthorlist}

\mlforastroaffiliation{yyy}{Los Alamos National Laboratory, Los Alamos, New Mexico, USA}
\mlforastroaffiliation{comp}{NASA Goddard Space Flight Center, Greenbelt, MD 20771, USA}

\mlforastrocorrespondingauthor{Abu Bucker Siddik}{siddik@lanl.gov}

\mlforastrokeywords{Light Curve, Pulsar, Parameter Estimation, Transformer}

\vskip 0.3in
]



\printAffiliationsAndNotice{\mlforastroEqualContribution} 

\begin{abstract}
Probing properties of neutron stars from photometric observations of these objects helps us answer crucial questions at the forefront of multi-messenger astronomy, such as, what is behavior of highest density matter in extreme environments and what is the procedure of generation and evolution of magnetic fields in these astrophysical environments? However, uncertainties and degeneracies—where different parameter sets produce similar light curves—make this task challenging. We propose a deep learning framework for inferring pulsar parameters from observed light curves. Traditional deep learning models are not designed to produce multiple degenerate solutions for a given input. To address this, we introduce a custom loss function that incorporates a light curve emulator as a forward model, along with a dissimilarity loss that encourages the model to capture diverse, degenerate parameter sets for a given light curve. We further introduce a test-time optimization scheme that refines predicted parameters by minimizing the discrepancy between the observed light curve and those reconstructed by the forward model from predicted parameters during inference. The model is trained using a suite of state-of-the-art simulated pulsar light curves. Finally, we demonstrate that the parameter sets predicted by our approach reproduce light curves that are consistent with the  true observation.
\end{abstract}

\section{Introduction}
\label{Introduction}
\begin{figure}[htbp]
\begin{center}
\centerline{\includegraphics[width=0.8\columnwidth]{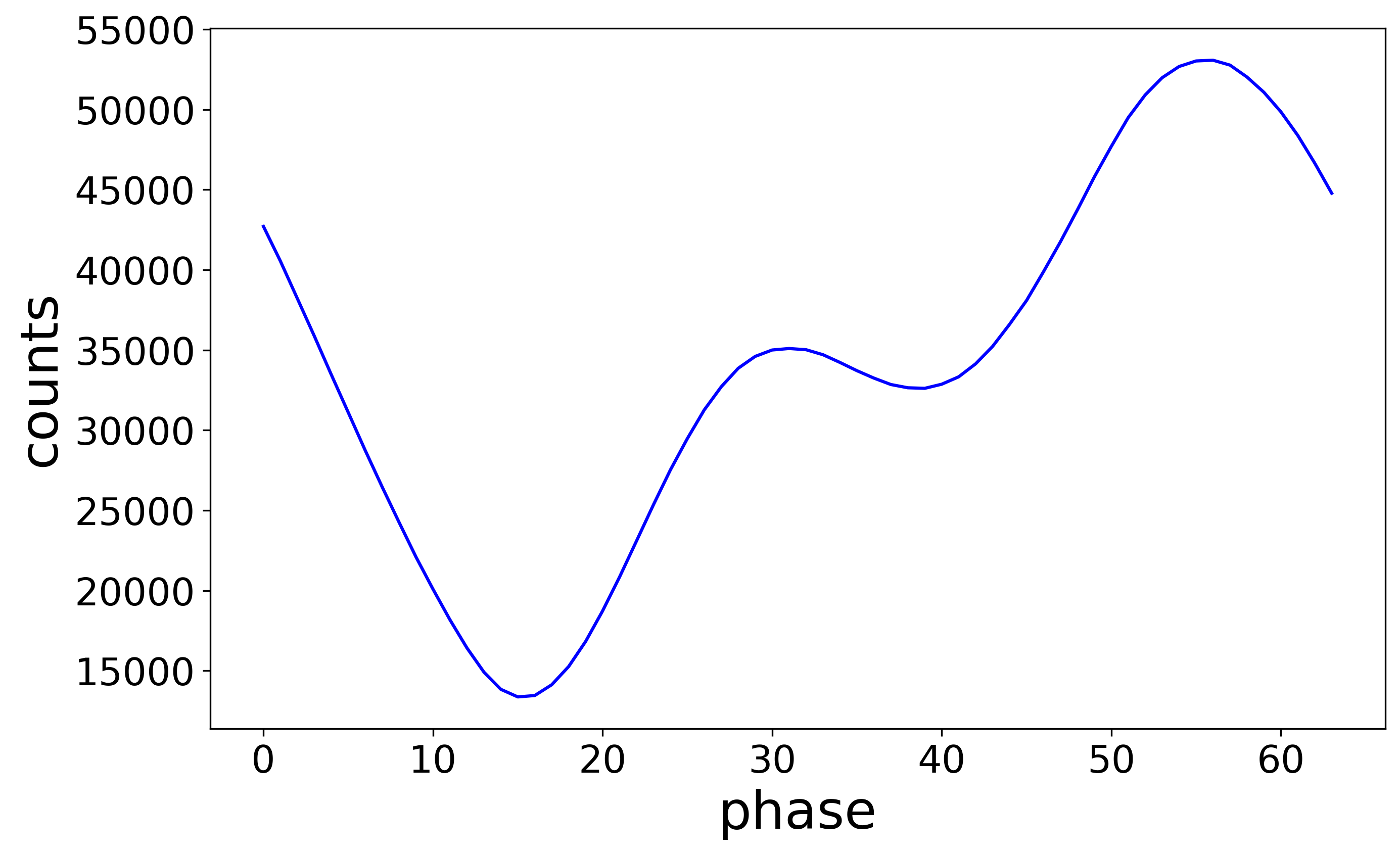}}
\caption{Simulated thermal light curve from a pulsar.}
\label{light-curve}
\end{center}
\end{figure}

Pulsars—rapidly rotating neutron stars—emit beams of electromagnetic radiation across all wavelengths, from $\gamma$-rays to radio  \cite{smith1977pulsars, sturrock1971model}. They emit radiation at regular intervals, making their light curves are periodic in nature, capturing variations in a pulsar's brightness over time. 
Figure \ref{light-curve} shows a simulated thermal light curve from a pulsar. 
In general, the relevant parameters that could be extracted from pulsar light curves include geometric ones (e.g., observer angle), aspects of the magnetic field structure, or more fundamental physical properties such as mass, radius (and thus the equation of state), or temperature \cite{benli2021constraining, brambilla2015testing, petri2021young, pierbattista2015light, venter2011modeling}. Estimation of these parameters play a key role in revealing the internal structure and dynamic processes of neutron stars \cite{cordes2004pulsars}. With numerous observations of pulsars, it is crucial to develop technologies that can extract optimal amount of information from these  observations while also completing the analyses in a reasonable amount of time. 
However, accurate prediction of parameters remains a challenging task due to the presence of uncertainties and degeneracies in the parameter estimation process. Degeneracy leads to multiple combinations of neutron star parameter sets generating nearly identical light curves, making it necessary to identify the variety of possible solutions of parameters given the observed light curve. Traditional pulsar parameter estimation methods, such as Markov Chain Monte Carlo (MCMC), are computationally  expensive—requiring months or even a year for analyzing a single light curve \cite{olmschenk2025pioneering}.

In this work, we present a novel deep learning framework for pulsar parameter estimation that explicitly accounts for uncertainties and degeneracies inherent in light curve data. Deep learning does not naturally predict multiple solutions, and so our approach: (1) incorporates a custom loss function that leverages a light curve emulator as a forward model; (2) encourages the exploration of degenerate solutions for the same input light curve; and, (3) introduces a test-time optimization strategy that refines model predictions during inference. Taken together, these methods produce parameter estimates that fit the given light curve well, 
identifies diverse regions of the parameter space that maps on to the given light curve, and gives strong parameter estimation accuracy on previously unseen light curves.

As far as we are aware, \textit{this is the first study to employ deep learning to explicitly address parameter degeneracies in pulsar light curve modeling} without relying on computationally expensive MCMC-based approaches. Rather than recovering the full posterior distribution, our model is designed to capture high-likelihood regions of the parameter space that are consistent with the observed light curves. Our approach offers a substantial reduction in computational cost, making it well-suited for scalable or real-time applications.

\section{Background and relevant work}
\label{Background}

\subsection{Degeneracies in parameter estimation from light curves}
\label{Degeneracies}
As discussed in Section \ref{Introduction}, degeneracies pose a significant challenge in pulsar parameter estimation from light curves. A single observed light curve can correspond to tens, hundreds, or even thousands of distinct parameter sets that yield the same or nearly indistinguishable light curves. Although machine learning techniques have shown considerable promise in astronomical inference tasks \cite{bino2023predicting, shi2023stellar}, such degeneracies significantly limit their effectiveness—particularly for traditional models, which are typically designed for single-point predictions rather than multimodal inference.

\subsection{Gaussian mixture model}
\label{GMM}
 A Gaussian Mixture Model (GMM) with $K$ components represents the data distribution \( p(\theta) \) as a weighted combination of $K$ Gaussian distributions \cite{reynolds2009gaussian}:
\begin{equation}
p(\theta) = \sum_{k=1}^K \pi_k \cdot \mathcal{N}(\theta | \mu_k, \sigma_k^2)
\label{eq:gmm}
\end{equation}
where \( \pi_k \) is the mixing coefficient (weight) of the \( k \)-th component, with \( \sum_{k=1}^{K} \pi_k = 1 \). Each \( \mu_k \in \mathbb{R}^{n} \) is an $n$-dimensional mean vector, and \( \sigma_k^2 \in \mathbb{R}^{n} \) is a vector of variances (assuming diagonal covariance). \( \mathcal{N}(\theta \mid \mu_k, \sigma_k^2) \) denotes a multivariate Gaussian distribution with independent components. In this work, the \( k \)-th components across the GMM collectively corresponding to indistinguishable light curves.

\subsection{Relevant work}
\label{Relevant Work}
Limited work exists on pulsar parameter estimation from light curves. \citet{kalapotharakos2021multipolar} demonstrated degeneracies in pulsar parameters 
and employed MCMC in combination with a physical model to perform parameter inference. The physical model is used to generate templates of light curves that is fit with the observed light light curve in the MCMC architecture. \citet{olmschenk2025pioneering}
 replaced the computationally expensive physical model with a neural network emulator for generating light curve templates within the MCMC framework. However, as MCMC involves generation of templates a million times to get the final posteriors, both approaches remain computationally expensive and slow: the method by \citet{kalapotharakos2021multipolar} takes approximately one year, whereas the \citet{olmschenk2025pioneering} approach reduces this to about 24 hours for a given light curve.

\section{Methods}
\label{Methods}

\subsection{Methodology}
\label{Methodology}

\begin{figure*}[ht]
\centering
\centerline{\includegraphics[width=\textwidth]{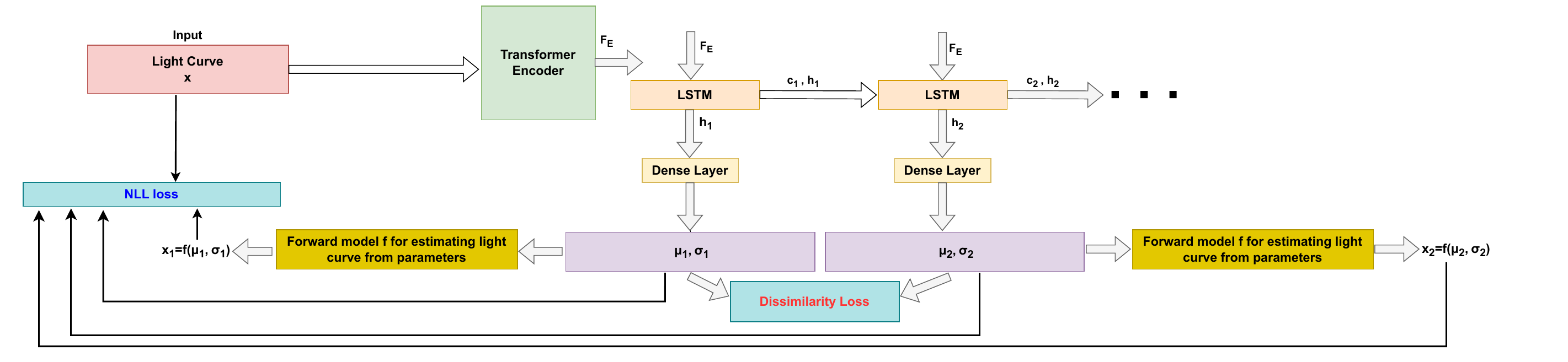}}
\caption{Block diagram of the proposed training methodology for pulsar parameter estimation from light curves. A transformer encoder first extracts key features $F_E$ from the input light curve $x$, which are then processed by a LSTM network that sequentially predicts the components of GMM $\{(\mu_1, \sigma_1), (\mu_2, \sigma_2), (\mu_3, \sigma_3), \ldots, (\mu_k, \sigma_k)\}$. To encourage diversity among the GMM components, a dissimilarity loss is employed. Each component $\theta_k=\mu_k, \sigma_k$ is passed through a forward model $f$ to generate corresponding light curve $\hat{\mathbf{x}}_k = f(\mu_k, \sigma_k)$. The negative log-likelihood (NLL) between the generated light curves and the input light curve is then computed to to guide the model toward better parameter estimation.}
\label{fig_methodology}
\end{figure*}

We propose a Transformer–LSTM architecture combined with a Gaussian Mixture Model (GMM) to estimate pulsar parameters from light curves. Figure \ref{fig_methodology} presents a block diagram of the proposed training methodology. 
 
\paragraph{Problem statement} Our goal is: Given light curve $x$, find a set of pulsar parameter values $\{\theta\}$ with high posterior likelihood $p(\mathbf{\theta}|x)$ s.t. $x = f(\mathbf{\theta})$ where $\mathbf{\theta}$ is a vector of 11 pulsar parameters and $f$ is a forward model. We approximate $p(\theta)$ with a GMM according to Eq~\ref{eq:gmm} $p(\theta) = \sum_{k=1}^K \pi_k \cdot \mathcal{N}(\theta | \mu_k, \sigma_k^2)$. In this study, we fix the number of mixture components to $K = 10$ and for simplicity, assume uniform mixing coefficients $\pi_k$, which are omitted from the GMM model.

\paragraph{Transformer encoder for feature extraction} To map a light curve $x$ to the mean and variance $(\mu, \sigma)$ of a normal distribution over parameters $\theta$, we employ a transformer encoder \cite{vaswani2017attention} to extract a latent feature representation $F_E$ from $x$. This feature vector is then passed to an LSTM network, which initially predicts the first component $(\mu_1, \sigma_1)$ of GMM using a dense layer corresponding to $\theta_1$.

\paragraph{Sequence of parameter predictions} To predict multiple sets of parameters, $\{\theta_1, \theta_2, \ldots, \theta_K\}$; the cell state $c_1$ and hidden state $h_1$ of initial prediction $(\mu_1, \sigma_1)$ of the LSTM network along with $F_E$ is fed into the same LSTM which predicts the subsequent components of the GMM $\{(\mu_2, \sigma_2), (\mu_3, \sigma_3), \ldots, (\mu_k, \sigma_k)\}$. To encourage these solutions to be different from each other, a dissimilarity loss is employed $\sum_{k=1}^{K-1}\sum_{k'=1}^{K-1} \mathcal{L}_{\textit{dis}}(\mu_k, \mu_{k'})$. 

To compute this dissimilarity, we use a radial kernel function $M(\theta, \theta')$ as a dissimilarity measure \cite{arora2023review}:

\begin{equation}
M(\theta, \theta')=\mathcal{L}_{\textit{dis}}(\mu_k, \mu_{k'})=\exp\left(-\frac{\|\theta - \theta'\|^2}{2c^2}\right)
\end{equation}
where $c$ is a hyperparameter that controls the kernel width.

\paragraph{Loss function based on forward model} Our training data contains only one parameter set per light curve; but the model should not be optimized to predict only that particular solution. Instead of using a standard loss function on the predicted $\hat{\theta} = \mu_k$ itself, we use a forward model $f$ \cite{olmschenk2025pioneering} which generates light curves $\hat{\mathbf{x}}_k = f(\mu_k, \sigma_k)$ and calculate the negative log-likelihood of the light curves from the predicted parameters against the given light curve $\sum_{k=1}^K \mathcal{L}_{\textit{NLL}}(\hat{\mathbf{x}}_k, x)$. Further details on the negative log-likelihood loss are provided in Appendix~\ref{appendix_A}.
 
 The overall training loss is $\mathcal{L} = \sum_{k=1}^K \mathcal{L}_{\textit{NLL}}(f(\mu_k, \sigma_k), x) + \gamma \sum_{k=1}^{K-1}\sum_{k'=1}^{K-1} \mathcal{L}_{\textit{dis}}(\mu_k, \mu_k')$, where $\gamma$ controls the relative contribution of the dissimilarity term. 

\paragraph{Test-time optimization} To enhance parameter estimation on previously unseen light curves, we apply a test-time optimization strategy during inference. Since the proposed framework relies solely on the input light curve and does not require ground-truth parameters, it can be optimized independently for each test instance. Specifically, the model is iteratively refined using the observed light curve by minimizing the mean squared error between the input light curve and those generated by the forward model using the predicted parameter sets. This procedure enables the model to adapt its predictions to the specific characteristics of each light curve, thereby improving the accuracy of the estimated parameters.

\subsection{Dataset for pulsar parameter estimation}
\label{Dataset}
We employ a comprehensive database of simulated X-ray light curves from pulsars, developed by Olmschenk et al. \cite{olmschenk2025pioneering}. Each light curve is associated with a unique set of physical parameters used for the simulation. 
Each parameter configuration is defined by 11 variables: the three Cartesian coordinates of the dipole location ($x_D$, $y_D$, $z_D$); the inclination angle ($\alpha_D$) and azimuthal angle ($\phi_D$) of the dipole moment; the three Cartesian coordinates of the quadrupole location ($x_Q$, $y_Q$, $z_Q$); the inclination angle ($\alpha_Q$) and azimuthal angle ($\phi_Q$) of the quadrupole moment; and the relative strength of the quadrupole component ($f_Q$). For this study, we utilize 10 million simulated light curves. 

\section{Results}
\label{Results}
\begin{figure*}[ht]
{\includegraphics[trim = 90 40 70 50, clip,width=.4\linewidth]{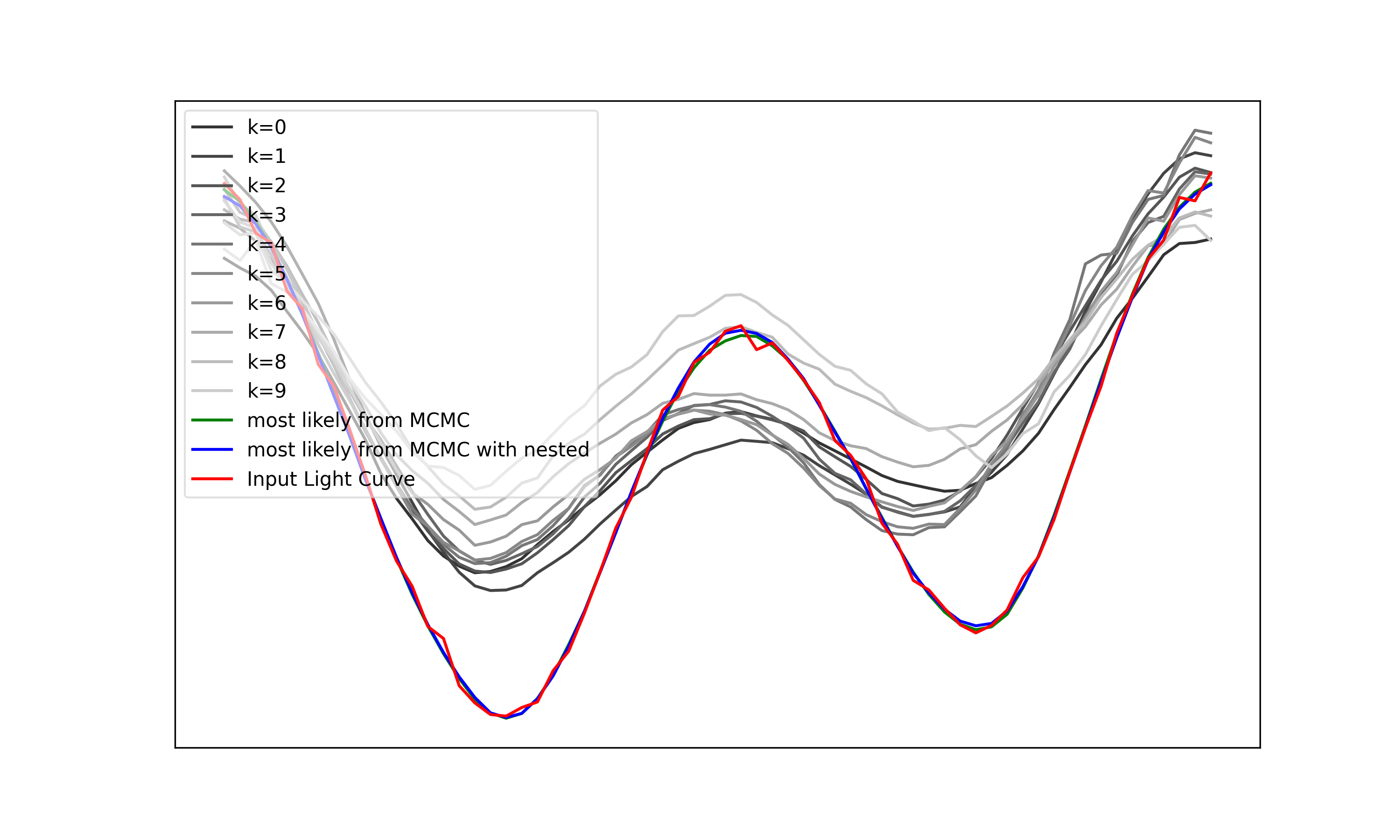}}\hfill
{\includegraphics[trim = 90 40 70 50, clip,width=.4\linewidth]{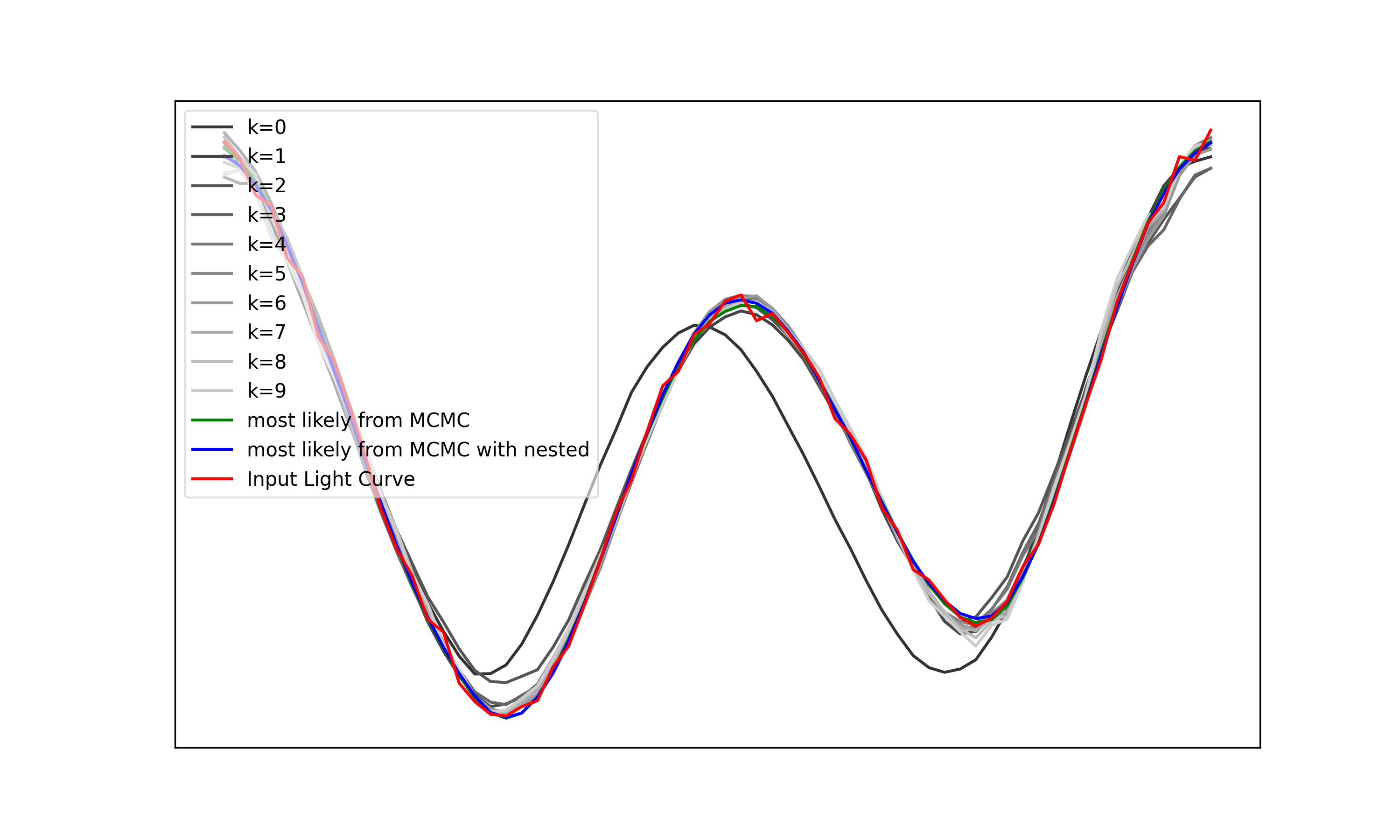}}
\caption{Comparison of predicted vs. true light curves for NICER's PSR J0030+0451 observation. The red curve is the real observation. The grey curves correspond to our predicted pulsar parameters upon analysis of this pulsar event. The green and blue curves correspond to most likely parameters produced by the traditional MCMC and MCMC with nested sampling methods respectively. The left and right panels show predictions from this work without and with  test-time optimization respectively. As can be seen, the test-time optimized predictions show strong alignment with the observation, and are comparable to those obtained using MCMC and MCMC with nested sampling.}
\label{fig:ttc}
\end{figure*}

\begin{table*}[ht]
\centering
\caption{Degenerate parameter sets for the light curve of pulsar PSR J0030+0451 with test-time optimization}
\label{compare}
\resizebox{\textwidth}{!}{%
\begin{tabular}{c|rrrrrrrrrrr}
\toprule
Case & $x_D$ & $y_D$ & $z_D$ & $\alpha_D$ & $\phi_D$ & $x_Q$ & $y_Q$ & $z_Q$ & $\alpha_Q$ & $\phi_Q$ & $Q/D$ \\
\midrule
parameter estimate for component $k=0$  & 0.2543 & -0.1974 &  0.2134 & 1.7291 & 2.7974 & 0.4851 & 0.2886 & -0.3096 & 2.0352 & 2.6625 & 6.4253 \\
parameter estimate for component $k=1$ & 0.4068 & -0.4145 & -0.0440 & 1.6552 & 2.4898 & 0.6214 & 0.3183 & -0.2811 & 1.8661 & 2.5606 & 9.7208\\
parameter estimate for component $k=2$  & 0.5204 & -0.2526 & -0.4009 & 1.6204 & 2.4512 & 0.6575 & 0.3720 & -0.2707 & 1.8258 & 2.6288 & 10.5740 \\
parameter estimate for component $k=3$ & 0.6108 & -0.0447 & -0.5904 & 1.5965 & 2.4643 & 0.6866 & 0.4193 & -0.2481 & 1.8258 & 2.6687 & 10.7498 \\
parameter estimate for component $k=4$  & 0.6472 &  0.1343 & -0.6943 & 1.5774 & 2.4932 & 0.6882 & 0.4561 & -0.2227 & 1.8266 & 2.6919 & 10.7489 \\
parameter estimate for component $k=5$  & 0.6398 &  0.2917 & -0.7584 & 1.5591 & 2.5306 & 0.6795 & 0.4871 & -0.2018 & 1.8221 & 2.7097 & 10.6697 \\
parameter estimate for component $k=6$  & 0.6024 &  0.4389 & -0.7988 & 1.5424 & 2.5719 & 0.6683 & 0.5123 & -0.1852 & 1.8148 & 2.7288 & 10.5388 \\
parameter estimate for component $k=7$  & 0.5466 &  0.5794 & -0.8219 & 1.5284 & 2.6136 & 0.6559 & 0.5312 & -0.1715 & 1.8073 & 2.7525 & 10.3759 \\
parameter estimate for component $k=8$  & 0.4764 &  0.7078 & -0.8326 & 1.5166 & 2.6527 & 0.6416 & 0.5449 & -0.1597 & 1.8015 & 2.7798 & 10.2041 \\
parameter estimate for component $k=9$ & 0.3920 &  0.8121 & -0.8337 & 1.5062 & 2.6868 & 0.6260 & 0.5549 & -0.1496 & 1.7981 & 2.8069 & 10.0481 \\
MCMC & -0.1662 & -0.0474 & -0.1081 & 1.2730 & 2.1249 & 0.4978 & 0.2616 & -0.0469 & 2.3353 & 2.4442 & 8.4039 \\
MCMC with nested sampling initialization & -0.1179 & 0.4110 & -0.1322 & 1.6895 & 2.1977 & 0.4637 & 0.2496 & -0.1769 & 0.7083 & 5.4385 & 4.7158 \\
\bottomrule
\end{tabular}%
}
\end{table*}

\begin{table}[ht]
\centering
\caption{Summary of MSE and MdNSE statistics between the true and predicted light curves over the testing dataset.}
\begin{tabular}{lcc}
\toprule
\textbf{Metric} & \textbf{Mean $\%$ error} & \textbf{Median $\%$ error} \\
\midrule
MSE    & 2.422 & 1.110 \\
MdNSE   & 1.286  & 0.523 \\
\bottomrule
\end{tabular}
\label{tab:mse_mdnse_stats}
\end{table}
 
To validate our predicted parameter sets $\hat{\theta}$ produced by testing our architecture with a given observed light curve, we feed the parameter solutions to the forward model $f$ and produce the corresponding predicted light curves $\hat{\mathbf{x}} = f(\hat{\theta}_{\mu_k})$. A strong agreement between the generated and given light curves indicates valid predictions. Figure \ref{fig:ttc} compares light curves generated from the predicted parameter sets with observed light curve from PSR J0030+0451, captured by NASA’s Neutron star Interior Composition Explorer (NICER) \cite{arzoumanian2014neutron}. For comparison, we also include light curves generated using the most likely parameter sets obtained via analysis of this event with MCMC and MCMC with nested sampling methods. Without test-time optimization, our model reproduces the overall shape of the input light curve but lacks precision. With test-time optimization, the predicted light curves closely align with the observation, achieving accuracy comparable to MCMC-based methods. These results highlight the effectiveness of the proposed framework for accurate pulsar parameter estimation from real observational data. Table \ref{compare} presents ten predicted parameter sets produced by our framework, alongside the most likely parameter sets obtained from MCMC and MCMC with nested sampling for the input light curve shown in Figure \ref{fig:ttc}. The MCMC-based results reveal evidence of multimodality in the parameter space—for example, in parameters such as $y_D, \alpha_Q$, where the most likely values differ significantly between the two methods. As shown, the proposed framework captures a similar range of multimodal values for some parameters and also identifies additional parameter sets that yield light curves closely matching the observed light curve, demonstrating its effectiveness in exploring the pulsar parameter space. The proposed framework also substantially reduces the computational cost of pulsar parameter estimation. Although it does not recover the full posterior distribution as traditional, computationally intensive MCMC methods do \cite{kalapotharakos2021multipolar, olmschenk2025pioneering}, it efficiently identifies high-likelihood regions of the parameter space that are consistent with the observed light curves, completing the analysis in approximately 4 minutes.

Table \ref{tab:mse_mdnse_stats} presents the mean squared error (MSE) and median normalized squared error (MdNSE) \cite{olmschenk2025pioneering} statistics between the true and predicted light curves across 1000 samples from the testing dataset. Test-time optimization is applied to the testing light curves, and both MSE and MdNSE are computed to quantify the alignment between the predicted and true light curves. The statistics reported in Table \ref{tab:mse_mdnse_stats} provide strong evidence of our model’s predictive accuracy, robustness, and its ability to generalize effectively across diverse test samples. Additional analysis is provided in Appendix \ref{appendix_B}.

\section{Conclusion}
\label{Conclusion}
We present a deep learning framework for estimating pulsar parameters from observed light curves. This task is challenging due to uncertainties and degeneracies in the parameter space, where different sets of parameters can produce similar light curves. We propose a custom loss function that combines a light curve emulator with a dissimilarity term, enabling the model to learn diverse parameter sets consistent with the same observation. We also apply test-time optimization to refine predictions on unseen light curves. For the observation PSR J0030+0451, our framework predicts parameter sets maps to light curves closely matching the observed light curve. Our predicted light curves are also comparable to those produced using most likely parameter values from analysis of the obsrervation with traditionally used MCMC methods. In this work, we effectively identify high-likelihood regions in the parameter space for a given light curve. To approximate this pulsar parameter distribution, we use a GMM with only 10 components. While this enables efficient modeling, it may constrain the full exploration of the parameter space. This limitation will be addressed in a future work.

\noindent\textbf{Impact Statement} This paper presents work whose goal is to advance the field of Machine Learning. There are many potential societal consequences of our work, none which we feel must be specifically highlighted here.

\noindent\textbf{Acknowledgments} This work was supported by the NASA theoretical and computational astrophysics network project with grant number 22-TCAN22-0027. This research used resources provided by the Los Alamos National Laboratory Institutional Computing Program, which is supported by the U.S. Department of Energy National Nuclear Security Administration under Contract No. 89233218CNA000001.

\bibliography{camera_ready_ml_astro}
\bibliographystyle{icml2025}

\newpage
\appendix
\onecolumn
\section{Negative Log-Likelihood Loss of Light Curves}
\label{appendix_A}
The overall training loss for the proposed framework can be rewritten as

\begin{equation}
    \mathcal{L} = \sum_{k=1}^K \mathcal{L}_{\textit{NLL}}(f(\mu_{k,j}, \sigma_{k,j}), x) + \gamma \sum_{k=1}^{K-1}\sum_{k'=1}^{K-1} \mathcal{L}_{\textit{dis}}(\mu_{k,j}, \mu_{k',j}),
\end{equation}

where the index $j$ corresponds to the 11 pulsar parameters, and $k$ denotes the components of the Gaussian Mixture Model (GMM). The first term, $\sum_{k=1}^K \mathcal{L}_{\textit{NLL}}(f(\mu_{k,j}, \sigma_{k,j}), x)$, represents the negative log-likelihood loss between the input light curve $x$ and the light curves generated from each predicted parameter set $(\mu_{k,j}, \sigma_{k,j})$. The second term, $\sum_{k=1}^{K-1}\sum_{k'=1}^{K-1} \mathcal{L}_{\textit{dis}}(\mu_{k,j}, \mu_{k',j})$, is a dissimilarity loss that encourages diversity among the degenerate parameter set predictions. The hyperparameter $\gamma$ controls the contribution of the dissimilarity term.

\paragraph{Negative Log-Likelihood for One Component of GMM}
The negative log-likelihood loss for one GMM component $(\mu_{k,j}, \sigma_{k,j})$, predicted by the LSTM network, is defined as:

\begin{equation}
    \mathcal{L}_{\textit{NLL}}(\mu_{k,j}, \sigma_{k,j}, \theta_{m,j})=\frac{1}{2} \sum_{j=1}^{11} \left( \frac{(\theta_{m,j} - \mu_{k,j})^2}{\sigma_{k,j}^2} + \log(2\pi \sigma_{k,j}^2) \right)
\end{equation} 

where $\theta_{m,j}$ denotes the ground truth pulsar parameter vector.

Since the true degenerate parameter sets $\theta_{m,j}$ for a input light curve $x$ are not available during training, we employ a forward model $f$ \cite{olmschenk2025pioneering} to generate light curves for each GMM component. Specifically, for a component $k$, we generate three light curves:
\begin{equation}
   \hat{x}_k = f(\mu_{k,j}) , \hat{x}_{k+\sigma_{k,j}} = f(\mu_{k,j}+\sigma_{k,j}), \hat{x}_{k-\sigma_{k,j}} = f(\mu_{k,j}- \sigma_{k,j}) 
\end{equation}

We then define the negative log-likelihood loss in the light curve space as:

\begin{equation}
    \mathcal{L}_{\textit{NLL}}(f(\mu_{k,j}, \sigma_{k,j}), x)=\frac{1}{2} \sum_{j=1}^{11} \left( \frac{L_{mse}(f(\mu_{k,j}, \sigma_{k,j}), x)}{\sigma_{k,j}^2} + \log(2\pi \sigma_{k,j}^2) \right)
\end{equation} 

where the MSE-based likelihood is defined as:
\begin{equation}
  L_{mse}(f(\mu_{k,j}, \sigma_{k, j}), x)=\frac{1}{3} \left(L_{mse}(\hat{x}_k, x) + L_{mse}(\hat{x}_{k+\sigma_{k,j}}, x) + L_{mse}(\hat{x}_{k-\sigma_{k,j}}, x)  \right)
\end{equation}

Incorporating $\hat{x}_{k+\sigma_{k,j}}$ and $\hat{x}_{k-\sigma_{k,j}}$ helps the model better estimate $\sigma_{k, j}$ by providing sensitivity to local perturbations in parameter space. The variance-based terms $\sigma_{k,j}^2$ and $\log(2\pi \sigma_{k,j}^2)$ act as regularization factors that penalize overconfident or underconfident uncertainty predictions of the model.

\clearpage
\section{Experimental Validation}
\label{appendix_B}
\begin{figure}[ht]
{\includegraphics[trim = 90 40 70 50, clip,width=.4\linewidth]{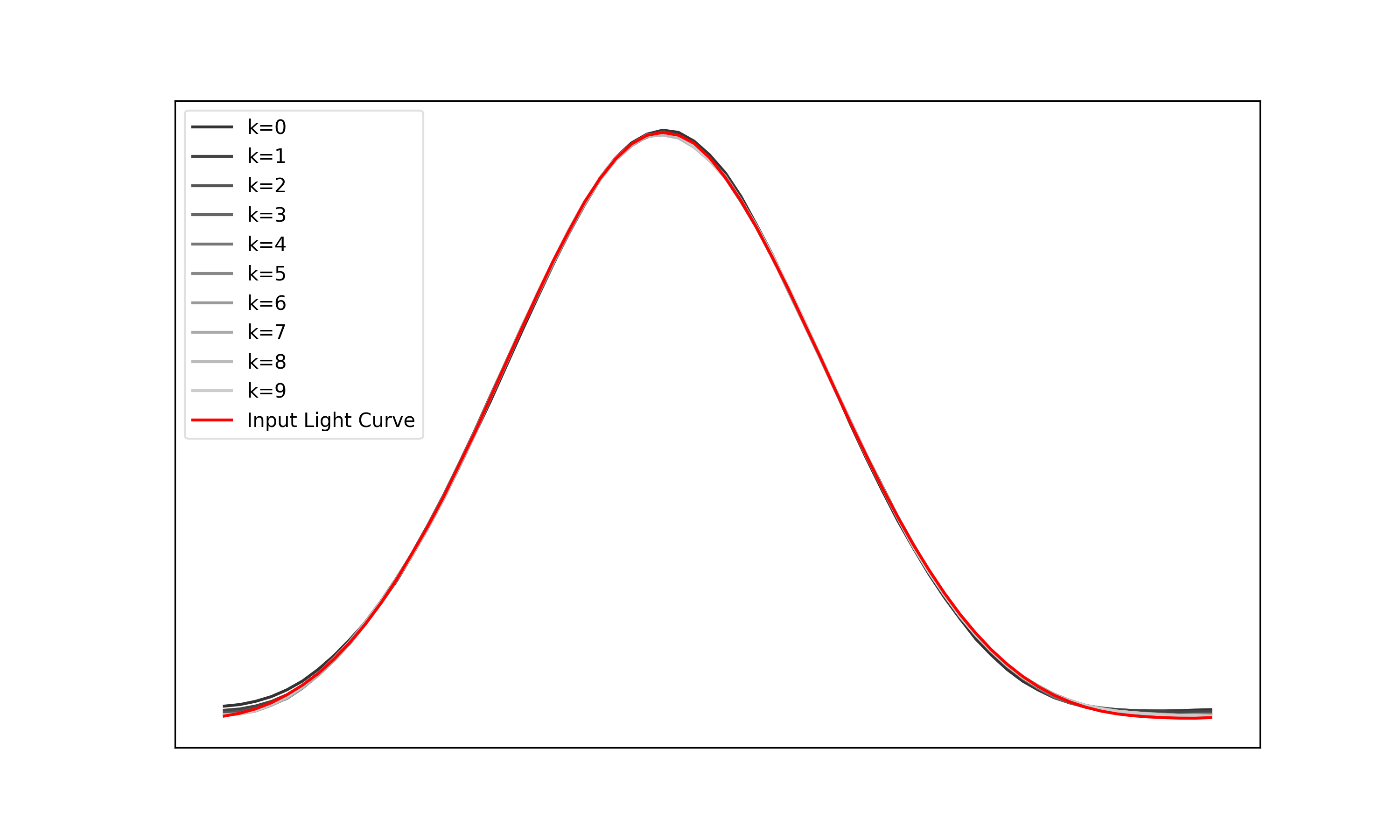}}\hfill
{\includegraphics[trim = 90 40 70 50, clip,width=.4\linewidth]{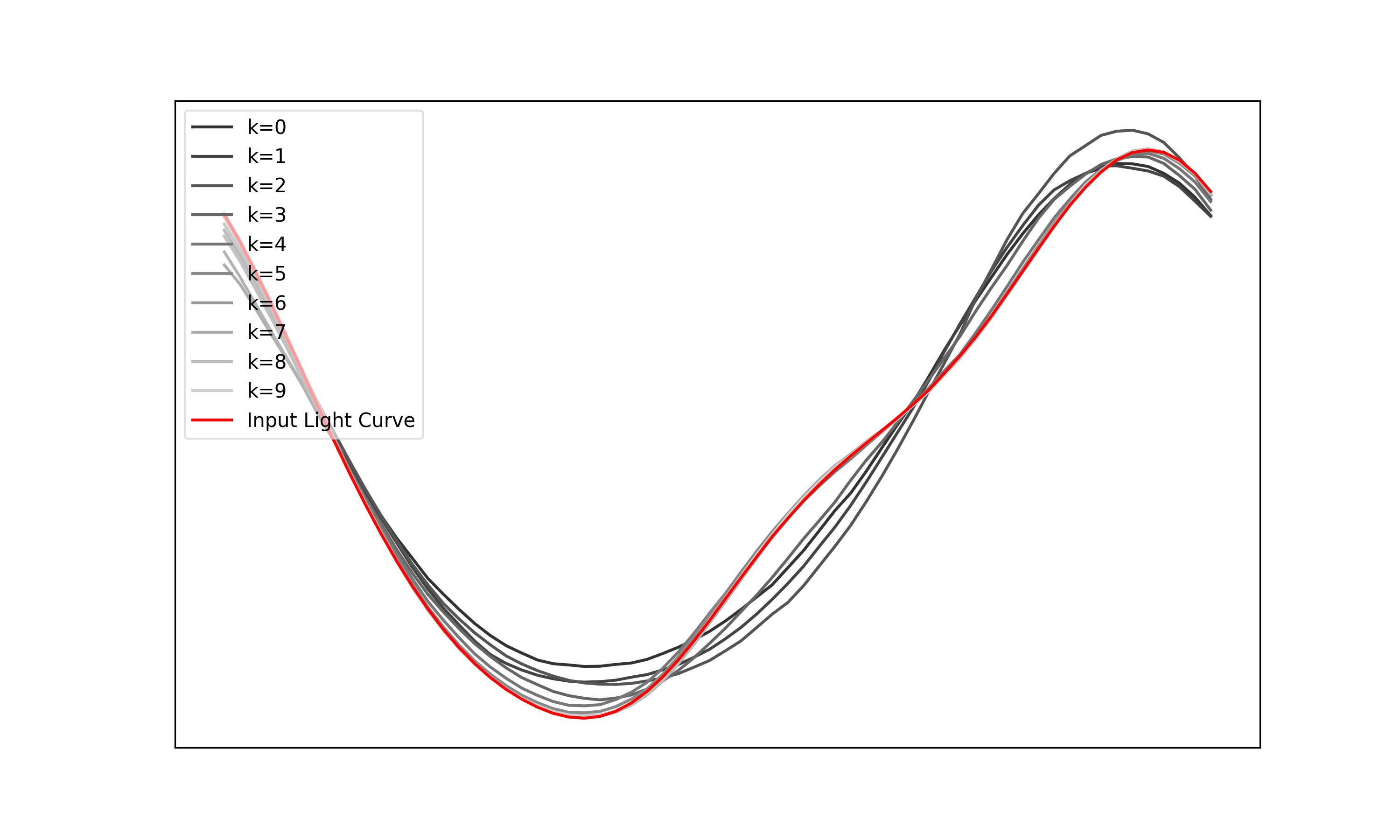}}
\caption{Comparison of predicted vs. true light curves for the testing data. The red curve represents the true light curve, while the grey curves show the predicted light curves based on our analysis of the pulsar event. The left and right panels illustrate the best-case and worst-case prediction scenarios, respectively, within the testing dataset.}
\label{fig:b-w}
\end{figure}

Figure \ref{fig:b-w} shows the best-case and worst-case parameter prediction scenarios within the testing dataset using the proposed methodology, based on the median normalized squared error (MdNSE) metric. In the best-case scenario, the predicted light curves align closely with the true light curve, yielding an MdNSE of 0.0012. In the worst-case scenario, the predicted light curves deviate noticeably from the true light curve, with an MdNSE of 0.15.

\section{Model training}
The model was trained using the Adam optimizer~\cite{diederik2014adam} for 15 epochs, with a step decay learning rate scheduler~\cite{ge2019step}. The initial learning rate was set to $10^{-5}$, with a step size of 5 epochs and a decay factor of 10. During inference, test-time optimization for a given observed light curve requires approximately 4 minutes to adapt the model's predictions to the specific input, using a single GPU.


\end{document}